\documentstyle[12pt]{article}  
\begin{document}  \bibliographystyle{unsrt}

\vbox{\vspace{15mm}}

\begin{center}

{\Large \bf Lorentz Group derivable from Polarization
Optics}\footnote{Presented at the 21st International Colloquium on
Group Theoretical Methods in Physics (Goslar, Germany, 1996)} \\[3mm]

D. Han\footnote{email: han@trmm.gsfc.nasa.gov}\\
{\it National Aeronautics and Space Administration, Goddard Space
Flight Center, Code 910.1, Greenbelt, Maryland 20771, U.S.A.}\\[5mm]

Y. S. Kim\footnote{email: kim@umdhep.umd.edu}   \\
{\it Department of Physics, University of Maryland, College Park,
Maryland 20742, U.S.A.} \\[5mm]

Marilyn E. Noz\footnote{email: noz@nucmed.med.nyu.edu}\\
{\it Department of Radiology, New York University, New York, New York
10016, U.S.A.}
\end{center}

\vspace{5mm}

\begin{abstract}
The Lorentz group is the fundamental language for space-time symmetries
of relativistic particles.  This group can these days be derived from
the symmetries observed in other branches of physics.  It is shown that
this group can be derived from optical filters.  The group O(2,1) is
appropriate for attenuation filters, while the O(3) group describes
phase-shift filters.  The combined operation leads to a two-by-two
representation of the six-parameter Lorentz group.  It is shown also
that the bilinear representation of this group is the natural language
for the polarization optics.
\end{abstract}

The Lorentz group serves useful purposes in many branches of physics.
In this note, we would like to show that the bilinear representation
of the six-parameter Lorentz group~\cite{barg47} is the natural language
for polarization of light waves.  In studying polarized light propagating
along the $z$ direction, the traditional approach is to consider the $x$
and $y$ components of the electric fields.  Their amplitude ratio and the
phase difference determine the degree of polarization.  Thus, we can
change the polarization either by adjusting the amplitudes, by changing
the relative phase shift, or both.  For convenience, we call the optical
device which changes amplitudes an ``attenuator,'' and the device which
changes the relative phase a ``phase shifter.''

Let us write the electric field vector as
\begin{equation}\label{expo1}
\pmatrix{E_{x} \cr E_{y}} =
\pmatrix{A \exp{\left\{i(kz - \omega t + \phi_{1})\right\}}  \cr
B \exp{\left\{i(kz - \omega t + \phi_{2})\right\}}} .
\end{equation}
where $A$ and $B$ are the amplitudes which are real and positive numbers,
and $\phi_{1}$ and $\phi_{2}$ are the phases of the $x$ and $y$
components respectively. This column matrix is called the Jones
vector~\cite{jones41,swind75}. The content of polarization is determined
by the ratio:
\begin{equation}
{E_{y}\over E_{x}} = \left({B\over A}\right) e^{i(\phi_{2} - \phi_{1})} .
\end{equation}
which can be written as one complex number:
\begin{equation}\label{ratio}
w = r e^{i\phi}
\end{equation}
with
$$
r = {B \over A} , \qquad \phi = \phi_{2} - \phi_{1} .
$$
The degree of polarization is measured by these two real numbers,
which are the amplitude ratio and the phase difference respectively.

The purpose of this paper is to discuss the transformation properties of
this complex number $w$.  The transformation takes place when the light
beam goes through an optical filter whose transmission properties are not
isotropic.  There are two transverse directions which are perpendicular to
each other.  The absorption coefficient in one transverse direction could
be different from the coefficient along the other direction.  Thus, there
is the ``polarization'' coordinate in which the absorption can be described
by
\begin{equation}\label{atten}
\pmatrix{e^{-\eta_{1}} & 0 \cr 0 & e^{-\eta_{2}}} =
e^{-(\eta_{1} + \eta_{2})/2} \pmatrix{e^{\eta/2} & 0 \cr 0 & e^{-\eta/2}}
\end{equation}
with $\eta = \eta_{2} - \eta_{1}$ .
This attenuation matrix tells us that the electric fields are attenuated at
two different rates.  The exponential factor $e^{-(\eta_{1} + \eta_{2})/2}$
reduces both components at the same rate and does not affect the degree of
polarization.  The effect of polarization is solely determined by the
squeeze matrix
\begin{equation}\label{sq1}
S(0, \eta) = \pmatrix{e^{\eta/2} & 0 \cr 0 & e^{-\eta/2}} .
\end{equation}
This type of mathematical operation is quite familiar to us from squeezed
states of light, if not from Lorentz boosts of spinors.  For convenience,
we call the above matrix an attenuator.

If the polarization coordinate is the same as the $xy$ coordinate where
the electric field components take the form of Eq.(\ref{expo1}), the
above attenuator is directly applicable to the column matrix of
Eq.(\ref{expo1}).  If the polarization coordinate is rotated by an angle
$\theta/2$, or by the matrix
\begin{equation}
R(\theta) = \pmatrix{\cos(\theta/2) & -\sin(\theta/2)
\cr \sin(\theta/2) & \cos(\theta/2)} ,
\end{equation}
then the squeeze matrix becomes
\begin{eqnarray}\label{sq2}
S(\theta, \eta) &=& R(\theta) S(0, \eta) R(-\theta) \\[2mm]  \nonumber
&=& \pmatrix{e^{\eta/2}\cos^{2}(\theta/2) + e^{-\eta/2}\sin^{2}(\theta/2) &
(e^{\eta/2} - e^{-\eta/2})\cos(\theta/2) \sin(\theta/2)
\cr (e^{\eta/2} - e^{-\eta/2})\cos(\theta/2) \sin(\theta/2)
& e^{-\eta/2}\cos^{2}(\theta/2) + e^{\eta/2}\sin^{2}(\theta/2)} .
\end{eqnarray}

Another basic element is the optical filter with two different values
of the index of refraction along the two orthogonal directions.  The
effect of this filter can be written as
\begin{equation}\label{phase}
\pmatrix{e^{i\lambda_{1}} & 0 \cr 0 & e^{i\lambda_{2}}}
= e^{-i(\lambda_{1} + \lambda_{2})/2}
\pmatrix{e^{-i\lambda/2} & 0 \cr 0 & e^{i\lambda/2}} ,
\end{equation}
with $\lambda = \lambda_{2} - \lambda_{1}$ .
In measurement processes, the overall phase factor
$e^{-i(\lambda_{1} + \lambda_{2})/2}$
cannot be detected, and can therefore be deleted.  The polarization
effect of the filter is solely determined by the matrix
\begin{equation}\label{shif1}
P(0, \lambda) = \pmatrix{e^{-i\lambda/2} & 0 \cr 0 & e^{i\lambda/2}} .
\end{equation}
This phase-shifter matrix appears like a rotation matrix around the
$z$ axis in the theory of rotation groups, but it plays a different
role in this paper.  We shall hereafter call this matrix a phase shifter.

Here also, if the polarization coordinate makes an angle $\theta$ with
the $xy$ coordinate system, the phase shifter takes the form
\begin{eqnarray}\label{shif2}
P(\theta, \lambda) &=& R(\theta) P(0, \lambda) R(-\theta)  \\[2mm] \nonumber
& = & \pmatrix{e^{-i\lambda/2}\cos^{2}(\theta/2) +
e^{i\lambda/2}\sin^{2}(\theta/2) &
(e^{-i\lambda/2} - e^{i\lambda/2})\cos(\theta/2) \sin(\theta/2) \cr
(e^{-i\lambda/2} - e^{i\lambda/2})\cos(\theta/2) \sin(\theta/2)
& e^{i\lambda/2}\cos^{2}(\theta/2) + e^{-i\lambda/2}\sin^{2}(\theta/2)} .
\end{eqnarray}

Since we are interested in repeated applications of these two different
kinds of matrices with different parameters, we shall work with the
generators of these transformations.  Let us introduce the Pauli spin
matrices of the form
\begin{equation}
\sigma_{1} = \pmatrix{1 & 0 \cr 0 & -1} , \quad
\sigma_{2} = \pmatrix{0 & 1 \cr 1 & 0} , \quad
\sigma_{3} = \pmatrix{0 & -i \cr i & 0} .
\end{equation}
These matrices are written in a different convention.  Here  $\sigma_{3}$
is imaginary, while $\sigma_{2}$ is imaginary in the traditional notation.
Also in this convention, we can construct three rotation generators
\begin{equation}
J_{i} = {1 \over 2} \sigma_{i} ,
\end{equation}
which satisfy the closed set of commutation relations
\begin{equation}\label{comm1}
\left[J_{i}, J_{j}\right] = i \epsilon_{ijk} J_{k} .
\end{equation}
We can also construct three boost generators
\begin{equation}
K_{i} = {i \over 2} \sigma_{i} ,
\end{equation}
which satisfy the commutation relations
\begin{equation}\label{comm2}
\left[K_{i}, K_{j}\right] = -i \epsilon_{ijk} J_{k} .
\end{equation}
The $K_{i}$ matrices alone do not form a closed set of commutation
relations, and the rotation generators $J_{i}$ are needed to form a
closed set:
\begin{equation}\label{comm3}
\left[J_{i}, K_{j}\right] = i \epsilon_{ijk} K_{k} .
\end{equation}

The six matrices $J_{i}$ and $K_{i}$ form a closed set of commutation
relations, and they are like the generators of the Lorentz group applicable
to the (3 + 1)-dimensional Minkowski space.  The group generated by the
above six matrices is called $SL(2,c)$ consisting of all two-by-two complex
matrices with unit determinant.

If we consider only the phase shifters, the mathematics is
basically repeated applications of $J_{1}$ and $J_{2}$, resulting in
applications also of $J_{3}$.  Thus, the phase-shift filters form an
$SU(2)$ or $O(3)$-like subgroup of the group $SL(2,c)$.  On the other
hand, if we consider only the attenuators, the mathematics
consists of repeated applications of $K_{1}$ and $J_{3}$, resulting in
applications also of $K_{2}$.  This is evident from the commutation
relation
\begin{equation}
\left[J_{3}, K_{1}\right] = i K_{2} .
\end{equation}
Indeed, $J_{3}, K_{1}$ and $K_{2}$ form a closed set of commutation
relations for the $Sp(2)$ or $O(2,1)$-like subgroup of
$SL(2,c)$~\cite{kitano89}.  This three-parameter subgroup has been
extensively discussed in connection with squeezed states of
light~\cite{knp91}.

If we use both the attenuators and phase shifters, the result is the
full $SL(2,c)$ group with six parameters.  The transformation matrix is
usually written as
\begin{equation}\label{abgd1}
L = \pmatrix{\alpha & \beta \cr \gamma & \delta} ,
\end{equation}
with the condition that its determinant be one:
$\alpha\delta - \gamma\beta$ = 1.  The repeated application of
two matrices of this kind results in
\begin{equation}\label{abgd2}
\pmatrix{\alpha_{2} & \beta_{2} \cr \gamma_{2} & \delta_{2}}
\pmatrix{\alpha_{1} & \beta_{1} \cr \gamma_{1} & \delta_{1}}
= \pmatrix{\alpha_{2}\alpha_{1} + \beta_{2}\gamma_{1} &
\alpha_{2}\beta_{1} + \beta_{2}\delta_{1} \cr
\gamma_{2}\alpha_{1} + \delta_{2}\gamma_{1} &
\gamma_{2}\beta_{1} + \delta_{2}\delta_{1}} .
\end{equation}
The most general form of the polarization transformation is the application
of this algebra to the column matrix of Eq.(\ref{expo1}).

We can obtain the same algebraic result by using the bilinear
transformation:
\begin{equation}\label{bilin1}
w' = {\delta w + \gamma \over \beta w + \alpha} .
\end{equation}
The repeated applications of these two transformations can be achieved
from
\begin{equation}\label{bilin2}
w_{1} = {\delta_{1} w + \gamma_{1} \over \beta_{1} w + \alpha_{1}} , \quad
w_{2} = {\delta_{2} w_{1} + \gamma_{2} \over \beta_{2} w_{1} + \alpha_{2}} .
\end{equation}
Then, it is possible to write $w_{2}$ as a function of $w$, and the result
is
\begin{equation}\label{bilin3}
w_{2} = {(\gamma_{2}\beta_{1} + \delta_{2}\delta_{1} ) w +
(\gamma_{2}\alpha_{1} + \delta_{2}\gamma_{1})
\over (\alpha_{2}\beta_{1} + \beta_{2}\delta_{1}) w +
(\alpha_{2}\alpha_{1} + \beta_{2}\gamma_{1})} .
\end{equation}
This is a reproduction of the algebra given in the matrix multiplication
of Eq.(\ref{abgd2}).  The form given in Eq.(\ref{bilin1}) is the bilinear
representation of the Lorentz group~\cite{barg47}.

Let us go back to physics.  If we apply the matrix $L$ of Eq.(\ref{abgd1})
to the column vector of Eq.(\ref{expo1}), then
\begin{equation}
\pmatrix{\alpha & \beta \cr \gamma & \delta} \pmatrix{E_{x} \cr E_{y}}
= \pmatrix{\alpha E_{x} + \beta E_{y} \cr \gamma E_{x} + \delta E_{y}} ,
\end{equation}
which gives
\begin{equation}
{E'_{y} \over E'_{x}} = {\gamma E_{x} + \delta E_{y} \over
\alpha E_{x} + \beta E_{y}} .
\end{equation}
In term of the physical quantity $w$ defined in Eq.(\ref{ratio}), this
formula becomes
\begin{equation}
w' = {\gamma + \delta w \over \alpha + \beta w} .
\end{equation}
This equation is identical to the bilinear form given in Eq.(\ref{bilin1}),
and the ratio $w$ can now be identified with the $w$ variable defined as
the parameter of the bilinear representation of the Lorentz group in the
same equation.  As we stated before, the purpose of this paper was to
derive the transformation property of this complex number, and the purpose
has now been achieved.  The bilinear representation of the Lorentz group
is clearly the language of optical filters and resulting polarizations.

This report is largely based on our recent paper on the bilinear
represention of the Lorentz group~\cite{hkn96}.

\end{document}